\documentclass[aps,prb,twocolumn,
    groupedaddress,superscriptaddress,
    amsfonts,amssymb,amsmath,floatfix,
    citeautoscript]{revtex4-1} % the two col format
\usepackage{graphicx} 
\usepackage[centering,hmargin=23mm,tmargin=29mm,bmargin=29mm]{geometry}
\usepackage{amsmath}
\usepackage{hyperref} 
\usepackage{multirow}
\usepackage{newtxtext}                                        
\usepackage{booktabs}
\usepackage{xcolor} 
\usepackage{hyperref}
\usepackage{float}
\usepackage[normalem]{ulem}
\usepackage[textwidth=1.5cm,textsize=footnotesize]{todonotes}
\hypersetup{colorlinks, 
    linkcolor={blue!75!black!80!yellow},
    citecolor={blue!75!black!80!yellow}, 
    urlcolor={blue!75!black!80!yellow}
    }
\usepackage[cmintegrals]{newtxmath}

\makeatletter
\renewcommand\@make@capt@title[2]{%
    \@ifx@empty\float@link{\@firstofone}{\expandafter\href\expandafter{\float@link}}%
    \sffamily{\textbf{#1}}\@caption@fignum@sep#2
}%

\makeatother

\newcommand{\HarvardSEAS}{John A. Paulson School of Engineering and Applied Sciences, Harvard University, Cambridge, MA, USA}

\begin{document} 

\author{Christina A. C. Garcia}\affiliation{\HarvardSEAS}
\author{Jennifer Coulter}\affiliation{\HarvardSEAS}
\author{Prineha Narang}\email{prineha@seas.harvard.edu}\affiliation{\HarvardSEAS}

\title{Optoelectronic Response of Type-I Weyl Semimetals TaAs and NbAs from First Principles}

\date{\today}

\begin{abstract} 

Weyl semimetals are materials with topologically nontrivial band structure both in the bulk and on the surface, hosting chiral nodes which are sinks and sources of Berry curvature. Weyl semimetals have been predicted, and recently measured, to exhibit large nonlinear optical responses. This discovery, along with their high mobilities, makes Weyl semimetals relevant to a broad spectrum of applications in optoelectronic, nanophotonic and quantum optical devices. While there is growing interest in understanding and characterizing the linear and nonlinear behavior of Weyl semimetals, an \emph{ab initio} calculation of the linear optical and optoelectronic responses at finite temperature remains largely unexplored. Here, we specifically address the temperature dependence of the linear optical response in type-I Weyl semimetals TaAs and NbAs. We evaluate from first principles the scattering lifetimes due to electron-phonon and electron-electron interaction and incorporate these lifetimes in evaluating an experimentally relevant frequency-, polarization- and temperature-dependent complex dielectric function for each semimetal. From these calculations we present linear optical conductivity predictions which agree well where experiment exists (for TaAs) and guide the way for future measurements of type-I Weyl semimetals. Importantly, we also examine the optical conductivity's dependence on the chemical potential, a crucial physical parameter which can be controlled experimentally and can elucidate the role of the Weyl nodes in optoelectronic response. Through this work, we present design principles for Weyl optoelectronic devices that use photogenerated carriers in type-I Weyl semimetals.

\end{abstract}

\maketitle

Weyl semimetals, one class of materials with topologically nontrivial electronic behavior, have generated considerable recent attention\cite{WeylDiracRMP,YanFelserRev2018,Hosur2013} for their novel responses to applied electric and magnetic fields. These materials exhibit linearly dispersive electronic band touchings in the bulk states of the crystal, which can be described by the Weyl equation\cite{Weyl1929} and appear in pairs of opposite chirality.\cite{Nielsen1981i,Nielsen1981ii} Weyl nodes are connected by characteristic Fermi arcs when projected onto the surface Brillouin zone; therefore Weyl semimetals are distinguished from other topological systems in having both unique bulk and surface states which are protected only by translational symmetry. Perhaps most promising, due to their separation of chirality, the Weyl nodes have diverging Berry connection, leading to predictions\cite{Chan2017,Konig2017,DeJuan2017,Zhang2018b,Zhang2018,Rostami2018} and observations\cite{Wu2017,Patankar2018,Osterhoudt2019,Ma2019,Sun2017} of strongly nonlinear optical responses in Weyl semimetals. In the context of the study of optoelectronic materials, nonlinearity has proven to be a powerful method for elucidating material properties, including the symmetries of electronic structure and their associated Berry curvature. Technologically, nonlinearity and harmonic generation is a key mechanism to generate high frequencies for electronics and optoelectronics, enable light-sources and lasers across broad wavelength ranges, and allow single and pair photon generation for quantum information science.\cite{basov2017towards, novotny2012principles, kauranen2012nonlinear} Further, the high mobilities\cite{Shekhar2015,Huang2015b,Ghimire2015,Zhang2015R,Luo2015,Hu2016,Arnold2016a,Du2016,Wang2016,Zhang2017,Sudesh2017} of Weyl semimetals, as well as their low carrier densities, make them especially well-suited for these device platforms.

Among different materials which have been proposed as Weyl semimetals, the most well-studied of these is TaAs\cite{Weng2015, Huang2015, Xu2015_TaAs, Lv2015_TaAs1, Lv2015_TaAs2, Yang2015_TaAs}, often in comparison to the structurally and compositionally related TaP\cite{Xu2015_TaP, Xu2016_TaP}, NbAs\cite{Xu2015_NbAs}, and NbP\cite{Xu2015_NbP, Souma2016_NbP}. A requisite condition for Weyl nodes to exist in a material is the breaking of inversion symmetry,\cite{Murakami2007} time-reversal symmetry,\cite{Wan2011} or both.\cite{Zyuzin2012a} Being of the first type, TaAs thus simultaneously satisfies the condition for a nonzero second-order nonlinear optical response. Realization of this concept has been observed in both experimental and theoretical studies of second-harmonic generation\cite{Wu2017,Patankar2018} and shift current\cite{Osterhoudt2019}, both second-order nonlinear responses. With the intense interest in the linear and nonlinear optoelectronic properties of TaAs and NbAs, we note a critical gap in the literature: \emph{ab initio} predictions of the optical behavior of these materials, in particular the frequency- and temperature-dependent dielectric response and optical conductivity. The temperature dependence of these quantities is of utmost importance for the application of these materials to realistic nanophotonic and optoelectronic devices\cite{giulianivignale, archambault2010quantum}. However, this has been largely inaccessible in prior work\cite{Grassano2018} because a first principles understanding of the electron and phonon states, as well as the electron-electron and electron-phonon coupling and scattering lifetimes incorporating the temperature dependence\cite{Ziman:1972} is essential to elucidate the complex dielectric function and transport coefficients in these materials.

In this \emph{Article}, we address the linear optical response in type-I Weyl semimetals TaAs and NbAs from first principles, evaluating an experimentally relevant frequency-, polarization- and temperature-dependent complex dielectric function for each material. We start by identifying the positions of the Weyl nodes and evaluate the lifetimes associated with different scattering in TaAs and NbAs from first-principles calculations, building towards a fully \emph{ab initio} complex dielectric function for each material. We then present the temperature-dependent optical conductivity, finding our calculations to be in excellent agreement with experimental observations of the real linear optical conductivity in TaAs; we expect our predictions for NbAs will spur optical measurements of this material and would show similar agreement with our calculations. Finally, we examine the chemical potential dependence of the optical conductivity in these semimetals and explore the relative contributions of the Weyl nodes to the linear optical response.

\begin{figure}
    \includegraphics[width=0.49\textwidth]{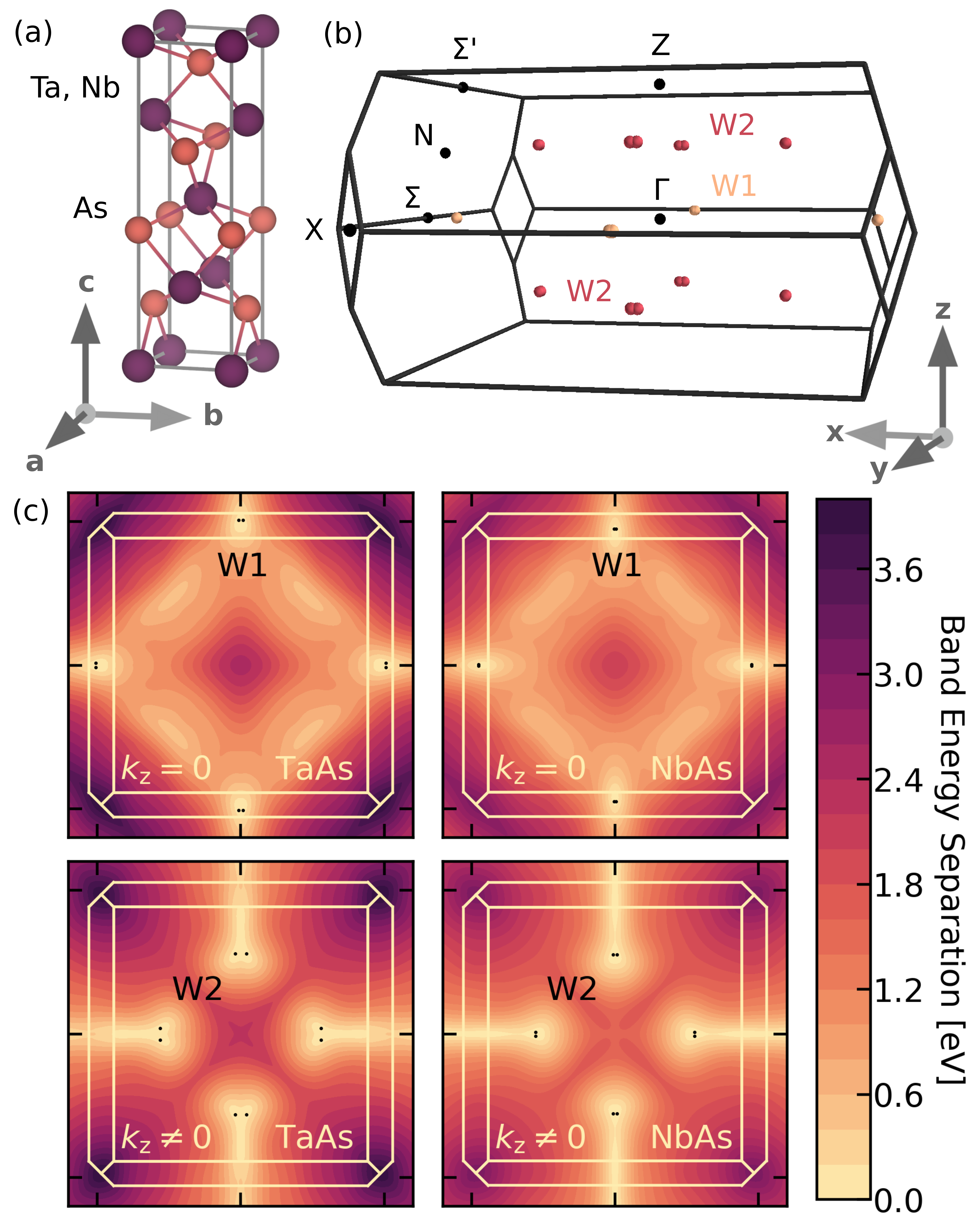}
    \caption{(a) Body-centered tetragonal crystal structure (\emph{I$4_1$md}) of TaAs and NbAs. (b) The first Brillouin zone of each semimetal hosts 12 pairs of Weyl nodes, denoted as W1 (in the $k_z=0$ plane) and W2 (above or below the $k_z=0$ plane). (c) The locations of these are shown more precisely with contour plots for TaAs and NbAs (left and right panels, respectively) denoting the energy separation of the two touching bands in the $k_z=0$ plane (top) and in the $k_z=\pm 0.578(2\pi/a)$ (TaAs) or $k_z=\pm 0.559(2\pi/a)$ (NbAs) plane (bottom), with the precise locations drawn on top in black.}
    \label{fig:structure}
\end{figure}

\textit{Structure and symmetry of transition metal monopnictides.} TaAs and NbAs crystallize in a body-centered tetragonal structure (\emph{I$4_1$md}, No. 109) as shown in Figure~\ref{fig:structure}(a). The first Brillouin zone of each material (Figure~\ref{fig:structure}(b)) contains 24 Weyl points which fall into two symmetrically nonequivalent categories. As presented in Figure~\ref{fig:structure}(c), there are 4 pairs which lie in the $k_z=0$ plane (denoted as W1, to follow convention\cite{WeylDiracRMP}) and 8 pairs which lie equidistantly above or below $k_z=0$ (W2). Each pair straddles either the $k_x=0$ or $k_y=0$ mirror plane. For TaAs and NbAs, respectively, we locate one such W1 at [$k_x$, $k_y$, $k_z$] = [0.008, 0.505, 0.000] and [0.003, 0.475, 0.000] at energies $-33$ meV and $-30$ meV, and W2 at [$k_x$, $k_y$, $k_z$] = [0.020, 0.280, 0.578] and [0.007, 0.277, 0.559] at energies $-37$ meV and $-20$ meV, in the fractional $k$-space coordinates of each standard conventional unit cell (Figure~\ref{fig:structure}(a)). We note that these Weyl point coordinates agree well with previous theoretical and experimental studies,\cite{Weng2015, Huang2015, Lee2015, Lv2015_TaAs2, Yang2015_TaAs, Xu2015_NbAs, Grassano2018} though their energetic positions relative to the Fermi energy (as well as the energy difference between W1 and W2) are different from other theoretical reports which lack consensus.\cite{Weng2015, Lee2015, Grassano2018}

\begin{figure*}
	\includegraphics[width=0.65\textwidth]{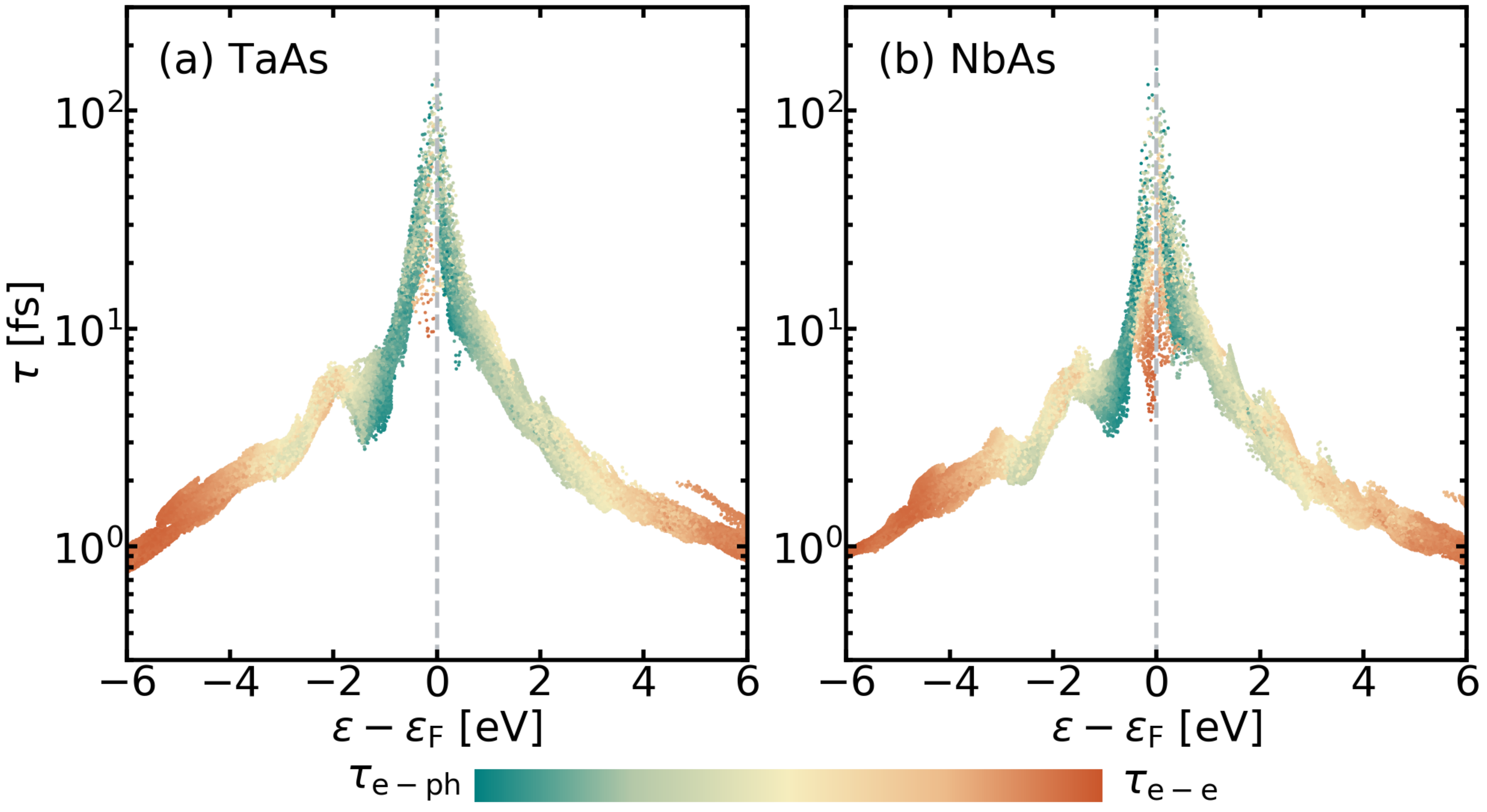}
	\caption{Room temperature ($T=298$ K) carrier scattering lifetimes near the Fermi level ($\varepsilon_\text{F}$) for (a) TaAs and (b) NbAs. The color bar indicates the relative contribution to scattering by electron-phonon (green) and electron-electron (orange) processes. As would be expected of a metal in a Fermi liquid theory regime, electron-phonon scattering dominates near $\varepsilon_\text{F}$, and electron-electron scattering dominates far away from $\varepsilon_\text{F}$. However, electron-electron scattering also dominates for some wavevectors \textbf{k} at energies very close to $\varepsilon_\text{F}$.}
	\label{fig:lifetimes}
\end{figure*}

\textit{Scattering lifetime evaluation.} First principles calculations of scattering lifetimes for TaAs and NbAs at room temperature ($T=298$ K) are shown as functions of energy in Figure~\ref{fig:lifetimes} at randomly sampled $k$-points throughout the first Brillouin zone. Here, we consider electron-electron and electron-phonon scattering, and the total scattering rate is given by Matthiessen's rule

\begin{equation}
    \tau_{\textbf{k}n}^{-1}=\left(\tau_{\textbf{k}n}^{\text{e-ph}}\right)^{-1}+\left(\tau_{\textbf{k}n}^{\text{e-e}}\right)^{-1}
\end{equation}

\noindent for electrons in band $n$ at wave vector \textbf{k}. We calculate the temperature-dependent electron-phonon scattering rate using Fermi's golden rule, as presented in Ref.~\citenum{ACSNanoBrown2016,Coulter2018,Ciccarino2018a,Narang2016a}:

\begin{multline}
    \left(\tau_{\textbf{k}n}^{\text{e-ph}}\right)^{-1}
    =\frac{2\pi}{\hbar}\int_{\text{BZ}}\frac{\Omega\text{d}\textbf{k}'}{(2\pi)^3}\sum_{n'\alpha\pm}
    \delta(\varepsilon_{\textbf{k}'n'}-\varepsilon_{\textbf{k}n}\mp\hbar\omega_{\textbf{k}'-\textbf{k},\alpha}) \\
    \times\left[n_{\textbf{k}'-\textbf{k},\alpha}+\frac{1}{2}\mp\left(\frac{1}{2}-f_{\textbf{k}'n'}\right)\right]
    \left|g_{\textbf{k}'n',\textbf{k}n}^{\textbf{k}'-\textbf{k},\alpha}\right|^2
    \label{eq:tau_e-ph}
\end{multline}

\noindent where $\Omega$ is the unit cell volume, $\varepsilon_{\textbf{k}n}$ and $f_{\textbf{k}n}=f(\varepsilon_{\textbf{k}n},T)$ are energies and Fermi occupations for an electron in band $n$ at wave vector \textbf{k}, $\hbar\omega_{\textbf{q}\alpha}$ and $n_{\textbf{q}\alpha}=n(\hbar\omega_{\textbf{q}\alpha},T)$ are energies and Bose occupations for a phonon with polarization index $\alpha$ at wave vector $\mathbf{q}$, and $g_{\textbf{k}'n',\textbf{k}n}^{\textbf{q},\alpha}$ is the electron-phonon coupling matrix element. By momentum conservation, only processes where $\textbf{q}=\textbf{k}'-\textbf{k}$ are allowed, and this is already substituted into (\ref{eq:tau_e-ph}). The electron-electron scattering rate is calculated from the imaginary part of the electron self-energy within the random phase approximation (RPA)\cite{ACSNanoBrown2016, Ladstadter2004} as:

\begin{multline}
    \left(\tau_{\textbf{k}n}^{\text{e-e}}\right)^{-1}
    =\frac{2\pi}{\hbar}\int_{\text{BZ}}\frac{\text{d}\textbf{k}'}{(2\pi)^3}\sum_{n'}\sum_{\textbf{GG}'}
    \Tilde{\rho}_{\textbf{k}'n',\textbf{k}n}(\textbf{G})\Tilde{\rho}^{*}_{\textbf{k}'n',\textbf{k}n}(\textbf{G}') \\
    \times\frac{1}{\pi}\operatorname{Im}\left[\frac{4\pi e^2}{|\textbf{k}'-\textbf{k}+\textbf{G}|^2}\epsilon^{-1}_{\textbf{GG}'}(\textbf{k}'-\textbf{k},
    \varepsilon_{\textbf{k}n}-\varepsilon_{\textbf{k}'n'})\right]
    \label{eq:tau_e-e}
\end{multline}

\noindent where $\Tilde{\rho}_{\textbf{k}'n',\textbf{k}n}$ are density matrices expressed in the plane-wave basis and $\epsilon_{\textbf{GG}'}$ is the RPA dielectric matrix for reciprocal lattice vectors \textbf{G} and \textbf{G}$'$. The temperature dependence of $(\tau_{\textbf{k}n}^{\text{e-e}})^{-1}$ is approximated with an analytical correction using Fermi liquid theory for electrons at temperature $T_e$:

\begin{equation}
    \left(\tau_{\textbf{k}n}^{\text{e-e}}\right)^{-1}(\varepsilon_{\textbf{k}n},T_e)\approx \frac{D_e}{\hbar}\left[(\varepsilon_{\textbf{k}n}-\varepsilon_\text{F})^2+(\pi k_BT_e)^2\right].
    \label{eqn:e-e_tempdep}
\end{equation}

\noindent We extract $D_e$ by fitting $(\tau_{\textbf{k}n}^{\text{e-e}})^{-1}$ from (\ref{eq:tau_e-e}) and then add $(D_e/\hbar)(\pi_BT_e)^2$ to (\ref{eq:tau_e-e}) for finite temperature results.\cite{PhysRevBBrown2016}

In Figure~\ref{fig:lifetimes}, we observe that the lifetimes for TaAs and NbAs look qualitatively very similar. As expected for a normal metal conforming to Fermi liquid theory, electron-electron scattering dominates far away from the Fermi energy ($\varepsilon_\text{F}$) due to a much larger phase space for such scattering events, reflected in the $(\varepsilon-\varepsilon_\text{F})^2$ dependence of (\ref{eqn:e-e_tempdep}). Moving closer to $\varepsilon_\text{F}$, electron-phonon scattering begins to dominate, again as in a Fermi liquid. This is where the largest carrier lifetimes occur due to the nearly-vanishing density of states at the Weyl point energies within tens of milli-electron volts of the Fermi level. Note, however, that close to $\varepsilon_\text{F}$, we also see the scattering rates at some points in reciprocal space dominated by electron-electron processes rather than by electron-phonon scattering, deviating from this conventional Fermi liquid theory picture, as we might expect at $\varepsilon_\text{F}$ for a vanishing density of states.

\begin{figure*}
    \includegraphics[width=0.96\textwidth]{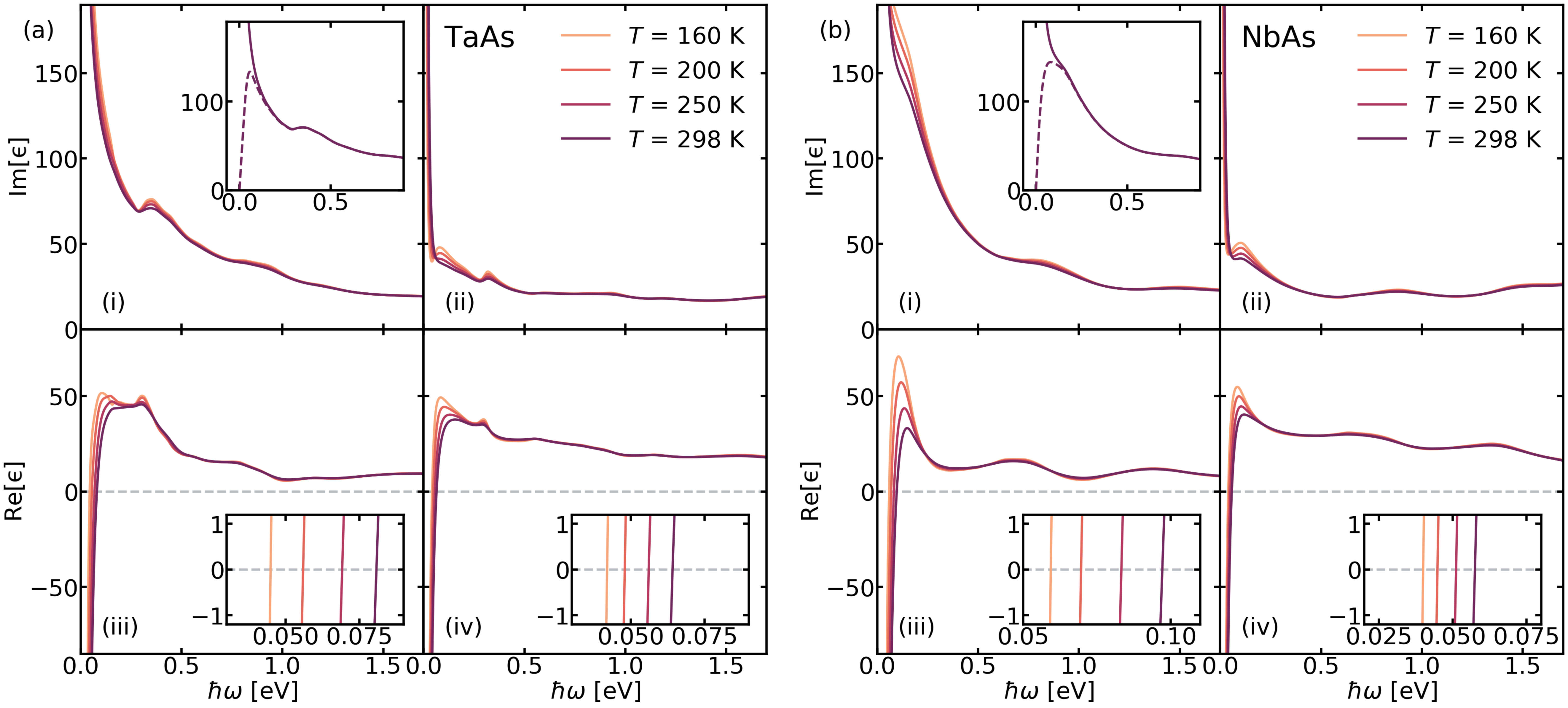}
    \caption{For (a) TaAs and (b) NbAs, the (i,ii) imaginary and (iii,iv) real parts of the complex dielectric function $\epsilon(\omega,T)$ polarized along $a$ (panels i,iii) and $c$ (panels ii,iv) crystallographic directions. The inset of each (i) shows the contribution from direct interband and phonon-assisted intraband transitions to $\operatorname{Im}[\epsilon]$ (dashed), dominated by direct transitions in the infrared regime. The insets of each (iii) and (iv) zoom in on where $\operatorname{Re}[\epsilon]$ changes sign for the $a$ and $c$ polarizations.}
    \label{fig:dielectric}
\end{figure*}

\textit{Dielectric function from first principles.} Next, we incorporate these lifetimes to study the optical responses of TaAs and NbAs, now computing the complex dielectric function of each as:\cite{ACSNanoBrown2016,Narang2016a}

\begin{equation}
    \bar{\epsilon}(\omega)=1+\mathrm{i}\frac{4\pi\bar{\sigma_0}}{\omega(1-\mathrm{i}\omega\tau_D)}+\bar{\epsilon}_{\text{d}}(\omega)+\bar{\epsilon}_{\text{ph}}(\omega)
    \label{eq:full-eps}
\end{equation}

\noindent where $\bar{\epsilon}_{\text{d}}(\omega)$ and $\bar{\epsilon}_{\text{ph}}(\omega)$ are the contributions from direct interband and phonon-assisted intraband excitations, respectively. The first term of (\ref{eq:full-eps}) accounts for the Drude response near $\varepsilon_\text{F}$, with $\tau_D$ being the average momentum relaxation time and $\sigma_0$ being the zero-frequency conductivity tensor. We calculate this conductivity using a linearized Boltzmann equation within a full-band relaxation-time approximation (RTA) as follows:

\begin{equation}
    \bar{\sigma_{0}}=\int_{\text{BZ}}\frac{e^2\text{d}\textbf{k}}{(2\pi)^3}\sum_{n}\frac{\partial f_{\textbf{k}n}}{\partial \varepsilon_{\textbf{k}n}}
    (v_{\textbf{k}n}\otimes v_{\textbf{k}n})\tau_{\textbf{k}n}^{\text{MR}}
    \label{eq:sigma0}
\end{equation}

\noindent where $v_{\textbf{k}n}$ are the band velocities and

\begin{equation}
\begin{split}
    \left(\tau_{\textbf{k}n}^{\text{MR}}\right)^{-1}
    &=\frac{2\pi}{\hbar}\int_{\text{BZ}}\frac{\Omega\text{d}\textbf{k}'}{(2\pi)^3}\sum_{n'\alpha\pm}
    \delta(\varepsilon_{\textbf{k}'n'}-\varepsilon_{\textbf{k}n}\mp\hbar\omega_{\textbf{k}'-\textbf{k},\alpha})\\
    & \times\left[n_{\textbf{k}'-\textbf{k},\alpha}+\frac{1}{2}\mp\left(\frac{1}{2}-f_{\textbf{k}'n'}\right)\right]
    \left|g_{\textbf{k}'n',\textbf{k}n}^{\textbf{k}'-\textbf{k},\alpha}\right|^2\\
    & \times\left(1-\frac{v_{\textbf{k}n}\cdot v_{\textbf{k}'n'}}{\left|v_{\textbf{k}n}\right|\left|v_{\textbf{k}'n'}\right|}\right)
\end{split} 
\label{eq:tau_MR}
\end{equation}

\noindent is identical to (\ref{eq:tau_e-ph}) except for an additional factor accounting for the change in momentum between final and initial states based on their relative scattering angle. Then the Fermi-surface averaged momentum relaxation time we use in (\ref{eq:full-eps}) is weighted by the Fermi occupation and square of the velocity:

\begin{equation}
    \tau_D=\frac{\int_{\text{BZ}}\frac{\text{d}\textbf{k}}{(2\pi)^3}
    \sum_{n}\frac{\partial f_{\textbf{k}n}}{\partial \varepsilon_{\textbf{k}n}}\left|v_{\textbf{k}n}\right|^2\tau_{\textbf{k}n}^{\text{MR}}}
    {\int_{\text{BZ}}\frac{\text{d}\textbf{k}}{(2\pi)^3}\sum_{n}\frac{\partial f_{\textbf{k}n}}{\partial \varepsilon_{\textbf{k}n}}\left|v_{\textbf{k}n}\right|^2}.
\label{eq:tau_D}
\end{equation}

In addition to the above components of the Drude contribution to the frequency dependent dielectric function, we also explicitly calculate the imaginary components of the dielectric function due to direct and phonon-assisted electronic transitions, as in our prior works.\cite{ACSNanoBrown2016,PhysRevBBrown2016, narang2017effects, papadakis2017ultralight} The corresponding expressions are Equation~\ref{eq:eps-direct} and \ref{eq:eps-phonon} in the Computational Methods. We then also predict the real part of the dielectric function \emph{via} the Kramers-Kronig relation. Importantly, the temperature dependence of Equation~\ref{eq:full-eps} is embedded primarily in Fermi (electron) and Bose (phonon) occupations which have explicit roles in each component of Equation~\ref{eq:full-eps} as well as implicit roles in renormalization involving the calculated scattering lifetimes.

\begin{figure}
    \includegraphics[width=0.5\textwidth]{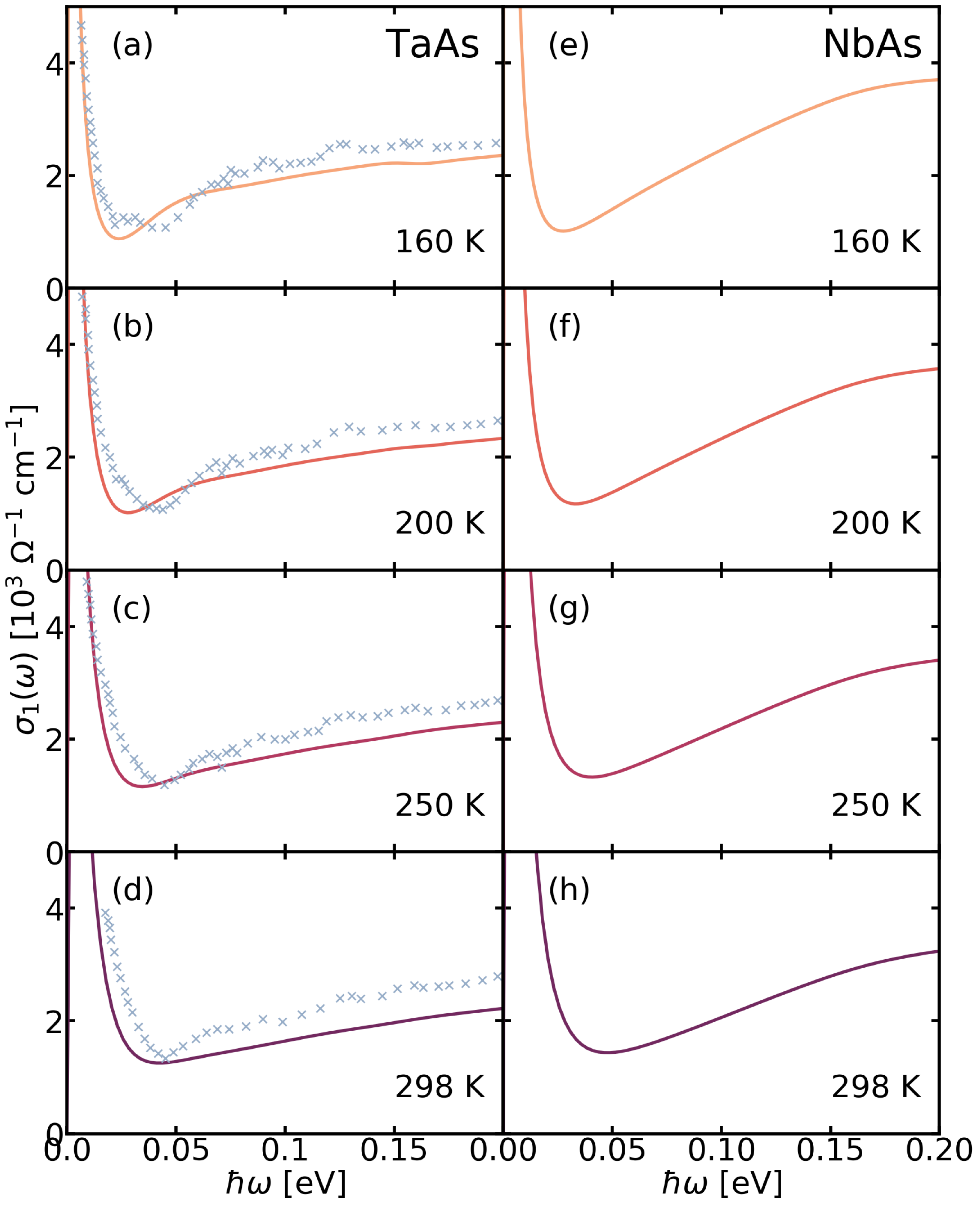}
    \caption{Real part of the optical conductivity tensor $\sigma_1$ as a function of frequency $\omega$ for both TaAs and NbAs polarized along the $a$ crystallographic direction at different temperatures. For TaAs, we also plot available experimental data from Kimura \emph{et al.}\cite{Kimura2017} on top for comparison.}
    \label{fig:sigma1}
\end{figure}

The real and imaginary parts of the complex dielectric function are presented for TaAs and NbAs as functions of frequency in Figure~\ref{fig:dielectric} for polarizations along the $a$ ($x,y$) and $c$ ($z$) crystallographic directions at several different temperatures ($T=$ 160, 200, 250 and 298 K) through the entire infrared (IR) range. As expected for a metal, starting at low frequencies (here, low mid-IR and below), the Drude component dominates, manifesting in a characteristic decay of $\operatorname{Im}[\epsilon]$ with increasing frequency. However, further into the mid-IR, a large contribution from direct interband transitions has almost the effect of broadening this decay of $\operatorname{Im}[\epsilon]$ with frequency. This contribution, added to the much smaller contribution from phonon-assisted intraband transitions, is shown as the dashed curve in the insets of Figure~\ref{fig:dielectric}(a,b)(i) for room temperature. Intuitively, we interpret the relatively small contribution from off-shell phonon-assisted transitions to result from the small interband threshold for TaAs and NbAs, as any such transition with energy exceeding this threshold is really a sequential process involving a direct transition, which we do not doubly count. This is in strong contrast to common plasmonic noble metals, for which this phonon-assisted contribution is significant.\cite{ACSNanoBrown2016}

The anisotropy of TaAs's and NbAs's optical response is most apparent in comparing $\operatorname{Im}[\epsilon]$ for $a$ and $c$ polarizations (Figure~\ref{fig:dielectric}(a,b)(i,ii)). Knowledge of this anisotropy would guide fabrication efforts and devices based on thin films as well as single-crystals of these materials. In general, loss in the linear regime of light polarized along the $c$ axis is lower than for along the $a$ or (equivalent) $b$ axes for the far and mid-IR. This is in part due to a narrower frequency band over which the Drude response dominates for the $c$ axis polarization compared to the $a$ axis for both materials. We also observe that the temperature dependence of the dielectric function is noticeably stronger in NbAs than in TaAs at low frequencies. This stems mainly from varying accessibility of direct interband transitions below $\sim$170 meV for the two materials; while TaAs exhibits a similar trend in the temperature dependence of the interband contribution, such transitions happen where the Drude response dominates and therefore are not as apparent in the overall dielectric function. We emphasize that the temperature dependence of the plasma frequency where $\operatorname{Re}[\epsilon]$ changes signs, highlighted in the insets of Figure~\ref{fig:dielectric}(a,b)(iii,iv), agrees well with previous experimental reports for TaAs.\cite{Xu2016,Kimura2017}

\textit{Linear optical conductivity prediction.} Next, we evaluate the optical conductivity of TaAs and NbAs. The real part of optical conductivity is related to the imaginary part of dielectric function as:

\begin{equation}
    \bar{\sigma_{1}}(\omega)=\epsilon_0\omega\operatorname{Im}[\bar{\epsilon}(\omega)].
\label{eq:sigma1}
\end{equation}

\begin{figure*}
    \includegraphics[width=0.65\textwidth]{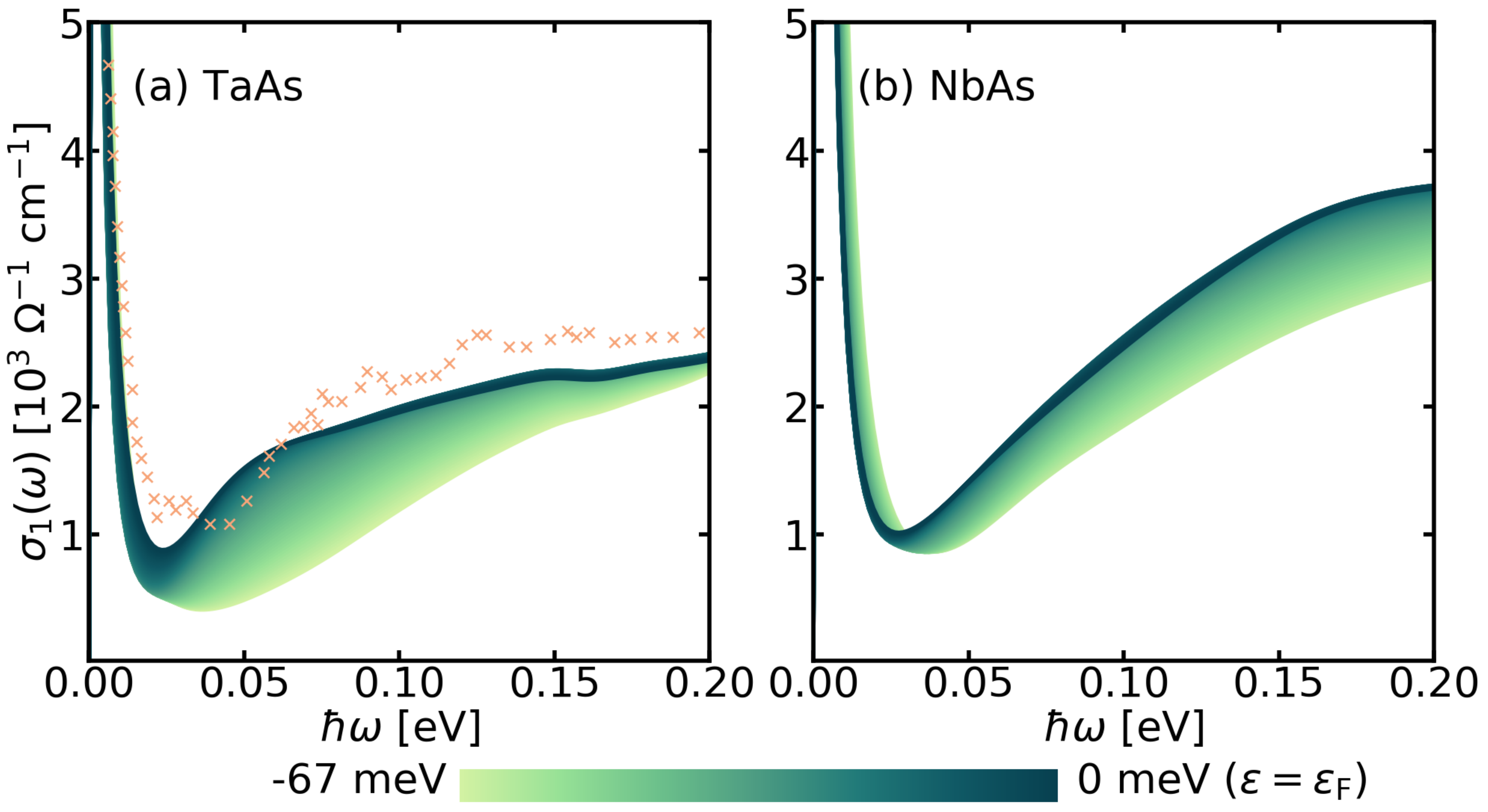}
    \caption{Real part of the optical conductivity tensor $\sigma_1$ as a function of frequency $\omega$ for both (a) TaAs and (b) NbAs polarized along the $a$ crystallographic direction and gated at different chemical potentials. The colorbar denotes the chemical potential shift referenced to the Fermi level. The same experimental data as in Figure~\ref{fig:sigma1} (Kimura \emph{et al.}\cite{Kimura2017}) is plotted for TaAs.}
    \label{fig:sigma1_dmu}
\end{figure*}

\noindent We plot in Figure~\ref{fig:sigma1} this frequency-dependent linear optical conductivity for the same four temperatures as in Figure~\ref{fig:dielectric} for light polarization along the $a$ crystallographic direction, up to 0.2 eV along the frequency axis. For TaAs, in Figure~\ref{fig:sigma1}(a)-(d), we overlay experimental data from Kimura \emph{et al.}\cite{Kimura2017} (grey `x's) for each temperature and find that our results are in excellent agreement both qualitatively and quantitatively. Our calculations deviate slightly from experiment in the diminishing of the Drude weight and corresponding onset of dominance by direct interband transitions, but this is readily explained by experimental uncertainty.

\textit{Effect of chemical potential variation.} Finally, we explore the role of the chemical potential in the optical response of TaAs and NbAs as another experimental parameter for optimizing these materials for optoelectronic devices. Specifically, we examine how the optical conductivity changes when the material is gated so that its chemical potential is shifted below $\varepsilon_{\rm F}$ to near the Weyl node energies. This is realized computationally by the inclusion of energetic adjustments to the chemical potential $\mu$ (otherwise equal to $\varepsilon_{\rm F}$) in the exponential arguments of the Fermi and Bose occupations of Equations~\ref{eq:sigma0}, \ref{eq:tau_MR}, and \ref{eq:tau_D}, as well as Equations~\ref{eq:eps-direct} and \ref{eq:eps-phonon}.

In Figure~\ref{fig:sigma1_dmu}, we present the real part of the linear optical conductivity for both TaAs and NbAs at 160 K (the lowest temperature considered in this work) with different chemical potential shifts below $\varepsilon_{\rm F}$, down to a minimum value of approximately twice the energy of the W1 Weyl node in each material (-33 meV in TaAs and -30 in NbAs). It is readily apparent that the slope and qualitative frequency dependence of the optical conductivity in NbAs does not change very much with chemical potential shift. This behavior of NbAs can be explained by the W1 nodes dominating the optical response in the photon energy range above $\sim$20 meV at a chemical potential $\mu=\varepsilon_{\rm F}$. The optical conductivity for energies above $\sim$20 meV in NbAs is only reduced once $\mu$ crosses below the energy of W1, where the phase space of allowed transitions near the W1 points diminishes, as less direct interband transitions and no phonon-assisted intraband transitions near the W1 points occur. In contrast, lowering the chemical potential in TaAs does change the qualitative behavior of its optical conductivity. When $\mu$ is tuned to be closer to the TaAs's Weyl node energies (W1 at -33 meV and W2 at -37 meV), the slope of $\sigma_1(\omega)$ near 20-50 meV becomes noticeably less steep, with this linear behavior extending farther out along the frequency axis with decreasing chemical potential. This suggests that transitions near the Weyl nodes dominate at $\mu=\varepsilon_{\rm F}$ only in this range from $\sim$20-50 meV and that with decreasing $\mu$, this frequency band widens while the optical conductivity decreases as in NbAs.

Importantly, these results place limitations on the common use of linear-in-$\omega$ behavior of $\sigma_1(\omega)$ as a clear optical diagnostic of Weyl physics, as the frequency range in which we might experimentally observe such behavior can be quite narrow and quantitatively dependent on the chemical potential of the sample. Further, we note that this frequency regime where we might target transitions near the Weyl nodes specifically for use in ``Weyltronic'' devices is one where these materials are very lossy, as seen in Figure~\ref{fig:dielectric}(a,b)(i,ii). 

\textit{Conclusions.} This work presents a comprehensive first principles study of the optoelectronic response of type-I Weyl semimetals TaAs and NbAs. By calculating the frequency-, polarization- and temperature-dependent complex dielectric function using a microscopic picture of the relevant electron-electron and electron-phonon interactions in each material, we establish a concrete link between Weyl physics, scattering processes and experimentally relevant optical quantities. Our work goes beyond the consideration of direct interband transitions and an estimated Drude contribution by including phonon-assisted intraband transitions and explicitly incorporating \textit{ab initio} quasiparticle lifetimes. We are thus able to predict the temperature-dependent linear optical conductivity without semi-empirical parameters and obtain excellent agreement with existing experimental observations for TaAs. Additionally, in examining the chemical potential dependence of the optical conductivity in these semimetals, we find the linear-in-$\omega$ behavior of $\sigma_1(\omega)$ to be a poor optical signature of Weyl node existence and urge experimentalists to use other methods in confirming the topological nature of new candidate materials.

The first principles methodology presented here is general and in future work could be used to calculate the response of newly discovered type-I and type-II Weyl semimetals. A similar study including the impact of chemical potential variation on nonlinear optical properties is an exciting topic of future work.

\textit{Computational Methods.} We first calculate the electronic structure of the two semimetals from first principles using density functional theory (DFT) as implemented by JDFTx\cite{JDFTx}. We use fully relativistic ultrasoft pseudopotentials\cite{DalCorsoPseudopotentials2014, RappeOptimized1990} for the PBEsol exchange correlation functional\cite{PBEsol} and obtain well converged results for a $3\times3\times3$ $k$-point mesh uniformly distributed across each 4-atom standard primitive unit cell. For TaAs and NbAs, we use plane-wave energy cutoffs of 22 and 24 Hartrees, respectively, and Marzari-Vanderbilt (``cold'') smearing\cite{Marzari1999} with a $0.001$ Hartree width. The phonon states are calculated using a frozen phonon approximation with a $3\times 3 \times 3$ supercell.

From the electronic states obtained through DFT, we construct a basis of maximally localized Wannier functions\cite{Souza2001} which we then use to interpolate all electron, phonon, and electron-phonon properties to much finer electron wave vector $\mathbf{k}$- and phonon wave vector $\mathbf{q}$ meshes.\cite{Giustino2007} We are thus able to very efficiently evaluate energies and matrix elements at arbitrary wave vectors for effective Monte Carlo Brillouin zone integration involved in calculating scattering rates and optical responses.

The imaginary component of the dielectric function of Equation~\ref{eq:full-eps} due to direct interband transitions is calculated as\cite{NatComSundararaman2014}

\begin{equation}
\begin{split}
    \boldsymbol{\lambda}\cdot\operatorname{Im}\bar{\epsilon}_{\text{d}}(\omega)\cdot\boldsymbol{\lambda}
    & =\frac{4\pi^2e^2}{\omega^2}\int_{\text{BZ}}\frac{\text{d}\textbf{k}}{(2\pi)^3}\sum_{n'n}(f_{\textbf{k}n}-f_{\textbf{k}n'})\\
    & \times\delta(\varepsilon_{\textbf{k}n'}-\varepsilon_{\textbf{k}n}-\hbar\omega)\left|\boldsymbol{\lambda}\cdot\left<\textbf{v}\right>_{n'n}^{\textbf{k}}\right|^2,
\label{eq:eps-direct}
\end{split}
\end{equation}

\noindent and similarly, the contribution from phonon-assisted electronic transitions is

\begin{equation}
\begin{split}
    \boldsymbol{\lambda}&\cdot\operatorname{Im}\bar{\epsilon}_{\text{ph}}(\omega)\cdot\boldsymbol{\lambda}
    =\frac{4\pi^2e^2}{\omega^2}\int_{\text{BZ}}\frac{\text{d}\textbf{k}\text{d}\textbf{k}'}{(2\pi)^6}\sum_{n'n\alpha\pm}(f_{\textbf{k}n}-f_{\textbf{k}'n'})\\
    &\times\left(n_{\textbf{k}'-\textbf{k},\alpha}+\frac{1}{2}\mp\frac{1}{2}\right)
    \delta(\varepsilon_{\textbf{k}'n'}-\varepsilon_{\textbf{k}n}-\hbar\omega\mp\hbar\omega_{\textbf{k}'-\textbf{k},\alpha})\\
    &\times\left|\boldsymbol{\lambda}\cdot\sum_{n_1}\left(\frac{g_{\textbf{k}'n',\textbf{k}n_1}^{\textbf{k}'-\textbf{k},\alpha}\left<\textbf{v}\right>_{n_1n}^{\textbf{k}}}
    {\varepsilon_{\textbf{k}n_1}-\varepsilon_{\textbf{k}n}-\hbar\omega+\mathrm{i}\eta}\right.\right.\\
    & \left.\left.+\frac{\left<\textbf{v}\right>_{n'n_1}^{\textbf{k}'}g_{\textbf{k}'n_1,\textbf{k}n}^{\textbf{k}'-\textbf{k},\alpha}}
    {\varepsilon_{\textbf{k}'n_1}-\varepsilon_{\textbf{k}n}\mp\hbar\omega_{\textbf{k}'-\textbf{k},\alpha}+\mathrm{i}\eta}\right)\right|^2.
\label{eq:eps-phonon}
\end{split}
\end{equation}

\vspace{1em}

\noindent Discussion of these contributions and details of how sequential processes above the interband threshold are eliminated from Equation~\ref{eq:eps-phonon} can be found in Ref.~\citenum{ACSNanoBrown2016}.

\textit{Acknowledgements.} This work used resources of the National Energy Research Scientific Computing Center, a DOE Office of Science User Facility, as well as resources at the Research Computing Group at Harvard University. Additional calculations were performed using resources from the Department of Defense High Performance Computing Modernization Program through the Army Research Office MURI grant on Ab-Initio Solid-State Quantum Materials: Design, Production, and Characterization at the Atomic Scale (18057522). This work was also partially supported by the STC Center for Integrated Quantum Materials, NSF Grant No. DMR-1231319. 

C.A.C.G. is supported by the NSF Graduate Research Fellowship Program under Grant No. DGE-1745303. J.C. recognizes the support of the DOE Computational Science Graduate Fellowship (CSGF) under Grant No. DE-FG02-97ER25308. P.N. is a Moore Inventor Fellow and a CIFAR Azrieli Global Scholar.

\bibliography{references}

\end{document}